# High yield production of graphene by liquid phase exfoliation of graphite

Yenny Hernandez[1*], Valeria Nicolosi[1*], Mustafa Lotya[1], Fiona M Blighe[1], Zhenyu Sun[1,2], Sukanta De[1,2], IT McGovern[1], Brendan Holland[1], Michelle Byrne[3], Yurii Gun'ko[2,3], John Boland[2,3], Peter Niraj[2,3], Georg Duesberg[2,3], Satheesh Krishnamurti[2,3], Robbie Goodhue[4], John Hutchison[5], Vittorio Scardaci[6], Andrea C. Ferrari[6], and Jonathan N Coleman[1,2**]

[1]*School of Physics, Trinity College Dublin, Dublin 2, Ireland*

[2]*Centre for Research on Adaptive Nanostructures and Nanodevices (CRANN), Trinity College Dublin, Dublin 2, Ireland*

[3]*School of Chemistry, Trinity College Dublin, Dublin 2, Ireland*

[4]*Department of Geology, School of Natural Sciences, Trinity College Dublin, Dublin 2, Ireland*

[5]*Department of Materials, University of Oxford, Parks Road, Oxford OX1 3PH, UK*

[6]*Engineering Department, University of Cambridge, 9 JJ Thomson Avenue, Cambridge CB3 0FA, U.K.*

*These authors contributed equally to this work.  **colemaj@tcd.ie

**Graphene is at the centre of nanotechnology research. In order to fully exploit its outstanding properties, a mass production method is necessary. Two main routes are possible: large-scale growth or large-scale exfoliation. Here, we demonstrate graphene dispersions with concentrations up to ~0.01 mg/ml by dispersion and exfoliation of graphite in organic solvents such as N-methyl-pyrrolidone. This occurs because the energy required to exfoliate graphene is balanced by the solvent-graphene interaction for solvents whose surface energy matches that of graphene. We confirm the presence of individual graphene sheets with yields of up to 12% by mass, using absorption spectroscopy, transmission electron microscopy and electron diffraction. The absence of defects or oxides is confirmed by X-ray photoelectron, infra-red and Raman spectroscopies. We can produce conductive, semi-transparent films and conductive composites. Solution processing of graphene opens up a whole range of potential large-scale applications from device or sensor fabrication to liquid phase chemistry.**

Graphene is one of the most exciting nano-materials due to the cascade of unique physical properties that have recently been demonstrated. For example, due to the details of its electronic structure, charge carriers in graphene behave as massless Dirac fermions[1]. Furthermore, novel effects such as an ambipolar field effect[2], room temperature quantum Hall effect[3], breakdown of the Born-Oppenheimer approximation[4] are observed. However, as was the case in the early days of nanotube and nanowire research, graphene at present still suffers from one problem, critical for its mass-scale exploitation: it cannot yet be made with high yield. The standard procedure used to make graphene is micromechanical cleavage[5]. This yields the best samples to date, with mobilities up to 200,000 $cm^2/Vs$.[6] However, single layers are a negligible fraction amongst large quantities of thin graphite flakes. Furthermore, it is difficult to see how to scale up this process to mass production. Alternatively, growth of graphene is also commonly achieved by annealing SiC substrates, but these samples are in fact composed of a multitude of domains, most of them sub-micrometer, and not spatially uniform in number, or in size over larger length scales[7]. A number of works have also reported graphene growth on metal substrates[8,9], but this would require the sample transfer to insulating substrates in order to make useful devices, either via mechanical transfer or, via solution processing.

Recently, a large number of papers have described the dispersion and exfoliation of graphene oxide (GO)[10-13]. This material consists of graphene-like sheets, chemically functionalised with compounds such as hydroxyls and epoxides, which stabilise the sheets in water[14]. However, this functionalisation results in considerable disruption of the electronic structure of the graphene. In fact GO is an insulator[15] rather than a semi-metal and is conceptually different from graphene. While the functionalities can be removed by reduction, large defect populations, which continue to disrupt the electronic properties, remain.[10,14] Thus, a non-covalent, solution-phase method to produce significant quantities of defect free, un-oxidised graphene is urgently required. In this paper we demonstrate such a method.

Here we show that high quality mono-layer graphene can be produced with large yields by non-chemical, solution phase exfoliation of graphite in certain organic solvents. This work builds upon over fifty years of study into chemical exfoliation of graphite[16]. Previously, intercalated graphite could be partially exfoliated by reactions involving the intercalant[17], through thermal shock[18] or by acid treatment of expandable graphite[19]. However, thus far, such methods give thin graphite sheets or graphene fragments[19] rather than large scale graphene mono-layers. The standard response to this problem has been the compromise of complete exfoliation of chemically modified forms of graphene such as graphene oxide or functionalised graphene [12,14,20]. However such materials are not graphene as they are insulators containing numerous structural defects[14,20], which cannot, so far, be fully removed by chemical treatment[14]. Our method results in high-quality, un-oxidised monolayer graphene at yields of ~1%. We show that the process can be improved to give yields of 12% with sediment recycling of the starting graphite mass. As a solution phase method, it is versatile and up-scalable and can be used to deposit graphene in a variety of environments and substrates not available using cleavage or growth methods. Furthermore, it can be used to produce graphene based composites or films, a key requirement for many applications, such as thin film transistors, conductive transparent electrodes for Indium Tin Oxide (ITO) replacement or for photovoltaics.

Recently, carbon nanotubes have been successfully exfoliated in a small number of solvents such N-methylpyrrolidone (NMP)[21-23]. Such exfoliation occurs because the strong interaction between solvent and nanotube sidewall means that the energetic penalty for exfoliation and subsequent solvation becomes small[22-24]. We suggest that similar effects may occur between these solvents and graphene. To test this we prepared a dispersion of sieved graphite powder (Aldrich product 332461, batch number 06106DE) in NMP by bath sonication. After sonication we obtained a grey liquid consisting of a homogenous phase and large numbers of macroscopic aggregates. As with nanotube dispersions[22,24], these aggregates could be removed by a mild centrifugation (CF), giving a homogenous dark dispersion. Such dispersions, prepared at different graphite concentrations are shown in Figure 1A. While moderate levels of sedimentation and aggregation occur within three weeks of centrifugation, the dispersions remain of high quality at least five months after preparation.

In order to find the concentration after CF, we filtered the graphite dispersion through polyvinylidene fluoride (PVDF) filters. Careful measurements of the filtered mass, accounting for residual solvent, gave the concentration of dispersed phase after centrifugation. This procedure was repeated for three other solvents known to successfully disperse nanotubes[25]: *N,N*-Dimethylacetamide (DMA), γ-Butyrolactone (GBL) and 1,3-Dimethyl-2-Imidazolidinone (DMEU). These dispersions were then characterised by UV-vis-IR absorption spectroscopy with the absorption co-efficient plotted versus wavelength in Figure 1B. The spectra are featureless in the visible-IR region as is expected from theory[26]. Each of these four dispersions were diluted a number of times and the absorption spectra recorded. The absorbance (660 nm) divided by cell length is plotted versus concentration in Figure 1C, showing Lambert-Beer behaviour for all solvents; $<\alpha_{660}>=2460$ Lg$^{-1}$m$^{-1}$.

Thus, it is clear that graphite can be dispersed in some solvents. As we will show, the graphite is almost completely exfoliated to multilayer structures with <5 layers in NMP, GBL and DMEU if not other solvents. In addition, significant quantities of individual monolayers are present. The question is what solvent properties lead to this exfoliation of graphite?

Such exfoliation can only occur if the net energetic cost is very small. In chemistry, this energy balance is expressed as the enthalpy of mixing (per unit volume) which we can approximately calculate in this case to be:

$$\frac{\Delta H_{Mix}}{V_{Mix}} \approx \frac{2}{T_{flake}}\left(\delta_G - \delta_{sol}\right)^2 \phi \qquad (1)$$

where $\delta_i = \sqrt{E_{Sur}^i}$, the square root of the component surface energy, $T_{flake}$ is the thickness of a graphene flake and φ is the graphene volume fraction. Reminiscent of the Hildebrand-Scratchard equation[27], this shows the enthalpy of mixing is dependent on the balance of graphene and solvent surface energies. For graphite the surface energy is defined as the energy per unit area required to overcome the van der Waals forces when peeling two sheets apart.

From equation 1, we expect a minimal energy cost of exfoliation for solvents whose surface energy matches that of graphene. To test this we dispersed graphite in a wide range of solvents. By measuring the optical absorbance (660nm) after CF and using the average absorption co-efficient to transform absorbance into concentration we can quantify the amount of graphite/graphene dispersed as a function of solvent surface energy (calculated from surface tension[28], see figure caption) as shown in figure 1D. As predicted, the dispersed concentration shows a strong peak for solvents with a surface energy very close to literature values of nanotube/graphite surface energy[29], ie ~70-80 mJ/m$^2$ (see arrow in figure 1D). Coupled with Equation 1, this strongly suggests that not only is the enthalpy of mixing for graphene dispersed in good solvents very close to zero, but the solvent-graphene interaction is van der Waals rather than covalent. In addition, it predicts that good solvents are characterised by surface tensions in the region of 40-50 mJ/m$^2$. Also, we can tell from these data that for the best solvent (DMEU), ~7% by mass of the original material remained after CF.

It is crucial to ascertain the exfoliation state of the material that remains dispersed after centrifugation. First we examine the state of the initial graphite powder. SEM studies (figure 2A) show the starting powder to consist of flakes of lateral size ~500μm and thickness <100μm. In comparison, the sediment separated after CF contains flakes, which are much smaller, with lateral size measured in tens of microns with thicknesses of a few microns (figure 2B). Clearly, sonication results in fragmentation of the initial flakes, with the largest removed by centrifugation. We note that, as the crystallite size in the starting powder was greater than 150 μm, the preparation procedure must result in tearing of the crystallites. This process may be similar to sonication induced fragmentation of carbon nanotubes[30].

We can investigate the state of the material remaining dispersed using TEM simply by dropping a small quantity of each dispersion onto holey carbon grids. Crucially, this technique is much simpler than that previously used to prepare graphene for TEM[31], which involved under-etching of graphene placed on a silicon substrate, and immediately shows one advantage of having graphene solutions. Shown in figure 2C-G are bright field TEM images of objects typically observed. We generally find three classes of object. The first, as shown in figure 2 C-E, are monolayer graphene. Secondly, in a number of cases we observe a single folded graphene layer (figure 2F). Thirdly we see bi-layer and multi-layer graphene (figure 2G). In all cases, these objects have lateral sizes of typically a few microns. In some cases the sheet edges tend to scroll and fold slightly (figure 2H). However, we have never observed large objects with thickness of more than a few layers. Thus we believe that, in these samples, graphite has been almost completely exfoliated to give monolayer and few-layer graphene. By analysing a large number of TEM images, paying close attention to the uniformity of the flake edges, we can generate flake thickness statistics (figure 2H). Thus, we can estimate the number fraction of monolayer graphene in NMP dispersions as 28% with a solution-phase monolayer mass fraction of ~12%, leading to an overall yield (mass of monolayers / starting graphite mass) of 0.8% (see table S2 for statistics). In fact, we also find that the sediment can be recycled to produce dispersions with number and mass fractions of monolayer graphene of ~18% and 7% respectively. This suggests the possibility of full sediment recycling and the eventual increase of the yield towards 7-12wt% (relative to the starting graphite mass).

A more definitive identification of graphene can be made by analysis of electron diffraction patterns[32]. As an example of this, in figure 3 A and B we show what appear to be a graphene monolayer and a graphene bi-layer respectively. Figure 3 B is particularly interesting as the right side of the flake consists of at least two layers while on the left side, a single mono-layer protrudes. Shown in figure 3 C is the normal incidence electron diffraction pattern of the flake in A. This pattern shows the typical six-fold symmetry expected for graphite/graphene[31,32] allowing us to label the peaks with the Miller-Bravais (hkil) indices. Shown in figure 3 D and E are normal incidence selected-area diffraction patterns for the flake in B, taken with beam positions close to the black and white dots respectively. This means we expect one pattern (D) to reflect monolayer graphene while the other (E) will reflect multi-layer graphene. In both cases we see a hexagonal pattern similar to that in C. The main difference between D and E is that for the multi-layer (E), the {2110} spots appear to be more intense relative to the {1100} spots. This is an important observation, as for graphene multi-layers with Bernal (AB) stacking, computational studies have shown that the intensity ratio, $I_{\{1100\}}/I_{\{2100\}}<1$, while for mono-layers $I_{\{1100\}}/I_{\{2100\}}>1$.[33] Virtually all the objects identified in all the images as multi-layers displayed $I_{\{1100\}}/I_{\{2100\}}<1$, demonstrating that Bernal stacking rather than AA stacking is predominant in these samples[33].

This identification of Bernal stacking in these thin multi-layers allows us to definitively differentiate mono-layer graphene from multi-layer graphene by inspection of the intensity ratio; $I_{\{1100\}}/I_{\{2100\}}$. To do this, we plot a line section through the (1-210)-(0-110)-(-1010)-(-2110) axis for the patterns in C, D and E in figures 3 F, G and H. In F and G we can clearly see that the inner peaks, (0-110) and (-1010), are more intense than the outer ones, (1-210) and (-2110), confirming that that both the flake in A, and the region marked by the black dot in B, are monolayer graphene. Conversely, H clearly shows inner peaks that are less intense than the outer ones confirming that the area around the white dot in B consists of more than one layer. Further confirmation of the presence of monolayer graphene can be found by measuring the diffraction peak intensity as a function of tilt angle as shown in section S2.8

We can use that fact that the ratio of the intensity of the {1100} to the {2110} peaks gives an unambiguous local identification of monolayer versus multilayer to provide information on the yield of monolayer graphene. We measured the diffraction pattern of 45 flakes before measuring the intensity ratio; $I_{\{1100\}}/I_{\{2110\}}$. These ratios are plotted as a histogram in figure 3I. We clearly see a bimodal distribution with peaks centred at $I_{\{1100\}}/I_{\{2110\}}=0.35$ and $I_{\{1100\}}/I_{\{2110\}}=1.5$ representing multilayer and monolayer graphene respectively. These results agree well with reported experimental intensity ratios of $I_{\{1100\}}/I_{\{2100\}}\sim0.4$ for a bilayer and $I_{\{1100\}}/I_{\{2100\}}\sim1.4$ for monolayer graphene.[32] Though this data suggests a yield of 51% monolayer graphene, this is certainly an over estimate as selected area electron diffraction can give mono-layer like patterns for multi-layers such as that in figure 3B, when the beam is incident on a protruding monolayer. Better statistics can be found by counting the number of monolayers per flake as shown in figure 2H. However, we can use electron diffraction to check the accuracy of our image analysis showing that we can reproducibly use it to identify monolayer graphene, thus confirming the results presented in figure 2H.

While figure 1D suggests a van der Waals type solvent-graphene interaction, it is crucial to definitively rule out any inadvertent basal-plane functionalisation, which could seriously alter the electronic structure. Raman spectra of thin films of graphite/graphene prepared by vacuum filtration onto alumina filters were also measured. The G line (~1580 cm$^{-1}$) and 2D line (~2700cm$^{-1}$) are clearly visible in all cases while the D peak (~1350cm$^{-1}$) is very weak or completely absent, except for the spectra of very small flakes (~0.5-2µm) which show edge effects[34]. This indicates that our process does not introduce significant structural defects[34], such as epoxides covalently bonded to the basal plane[14]. In addition we recorded Raman spectra for individual flakes deposited on marked TEM grids allowing us to identify monolayers, bilayers and multilayers from the shape of the 2D band, emphasising the quality of our exfoliation. The lack of pronounced D bands in these individual flakes demonstrates the lack of defects. Furthermore, we have used X-ray photoelectron spectroscopy and infra red spectroscopy to show the complete absence of oxidisation typically associated with GO[10,11]. These experiments show categorically that we can produce high-quality, un-oxidised graphene in solution with high yield.

We can briefly illustrate the potential of this method of graphene production by using it to make thin graphene films. Raman and SEM analyses show that these films consist predominately of thin graphene multilayers with less than 5 layers. X-ray photoelectron spectroscopy measurements show that these films have ~11wt% residual NMP after drying at room temperature at ~10$^{-3}$ mbar. This value remained unchanged after a subsequent vacuum anneal at 400C. Combustion analysis gives an NMP content of ~10wt% after room temperature drying (~10$^{-3}$ mbar) which can be reduced to <7wt% after annealing. These films have conductivities of ~6500 S/m, similar to reduced graphene oxide films[11] and optical transparencies of ~42%. We also demonstrate polystyrene-graphene composites in NMP at high volume fraction. We measured the conductivity of such composites as a function of graphene volume to scale from 45 S/m for a 70vol% film to 476 S/m for a film of graphene/graphite. These values are comparable to the most conductive polymer-nanotube composites[35] and significantly higher than those quoted for graphene-oxide based composites[12]. Finally, we have deposited graphene monolayers on SiO$_2$ surfaces via spray coating, demonstrating that this processing method can potentially be used to prepare samples for microelectronic applications.

In conclusion, we have demonstrated a scalable, high yield method to produce high-quality, un-oxidised graphene from powdered graphite. By using certain solvents, graphene can be dispersed at concentrations of up to 0.01 mg/ml. These dispersions can then be used to deposit flakes by spray coating, vacuum filtration, drop casting or by adding polymers can be turned into polymer-composite dispersions. We believe that this work opens up a whole new vista of potential applications from sensor or device applications to transparent electrodes and conductive composites.

Acknowledgements

We thank CRANN and SFI-RFP for financial support and Nacional de Grafite for supplying flake graphite. VN wishes to thank the EU project ESTEEM for facilitating access to the microscopy facilities in Oxford. ACF acknowledges funding from the Leverhulme Trust and the Royal Society.

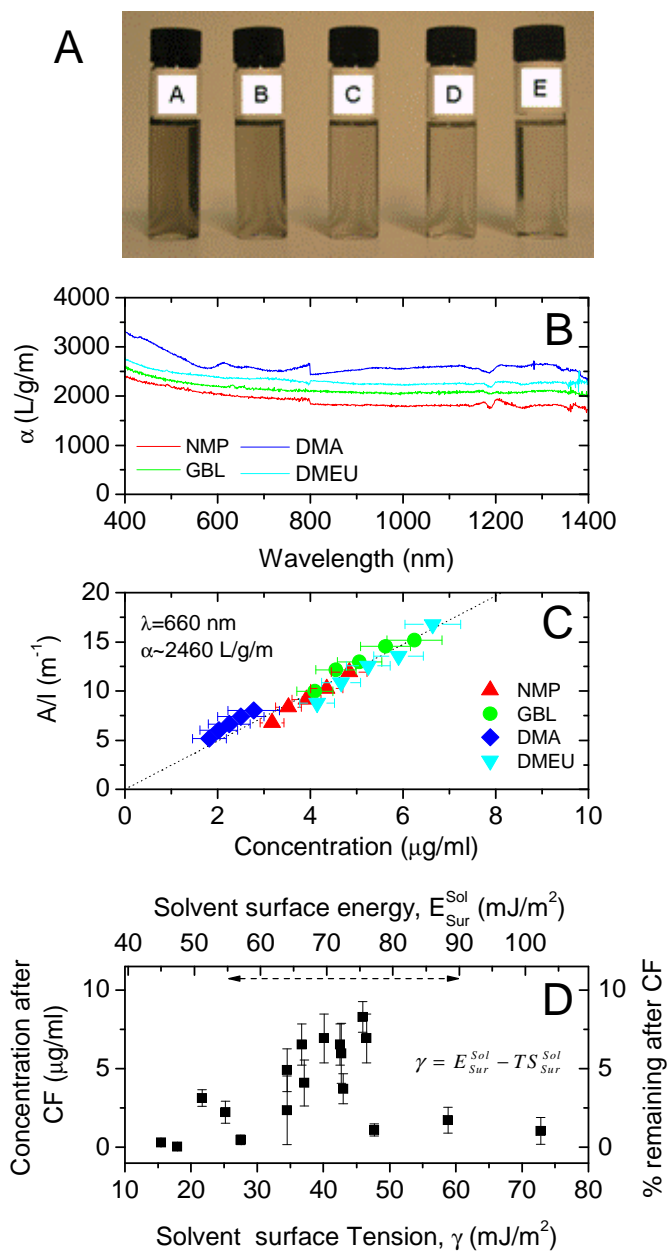

Figure 1 A) Dispersions of graphene in NMP, at a range of concentrations ranging from 6 – 4 µg/ml (A-E) after centrifugation. B) Absorption spectra for graphene dispersed in NMP, GBL, DMA and DMEU at concentrations from 3 to 9 µg/ml. C) Optical absorbance (660 nm) divided by cell length (A/l) as a function of concentration for graphene in the 4 solvents described above showing Lambert-Beer behaviour with an average absorption coefficient of $<\alpha_{660}>$=2240 L/g/m. D) Graphene concentration measured after centrifugation for a range of solvents plotted versus solvent surface tension. This data was converted from absorbance using $A_{660}/l=<\alpha_{660}>C$ with $<\alpha_{660}>$=2240 L/g/m. The original concentration, before

centrifugation, was 0.1 mg/ml. Shown on the right axis is the percentage of material remaining after CF. On the top axis, the surface tension has been transformed into surface energy using a universal value for surface entropy of ~0.1 mJ/m$^2$K. The horizontal arrow shows the approximate range of the reported literature values for the surface energy of graphite [ref 31].

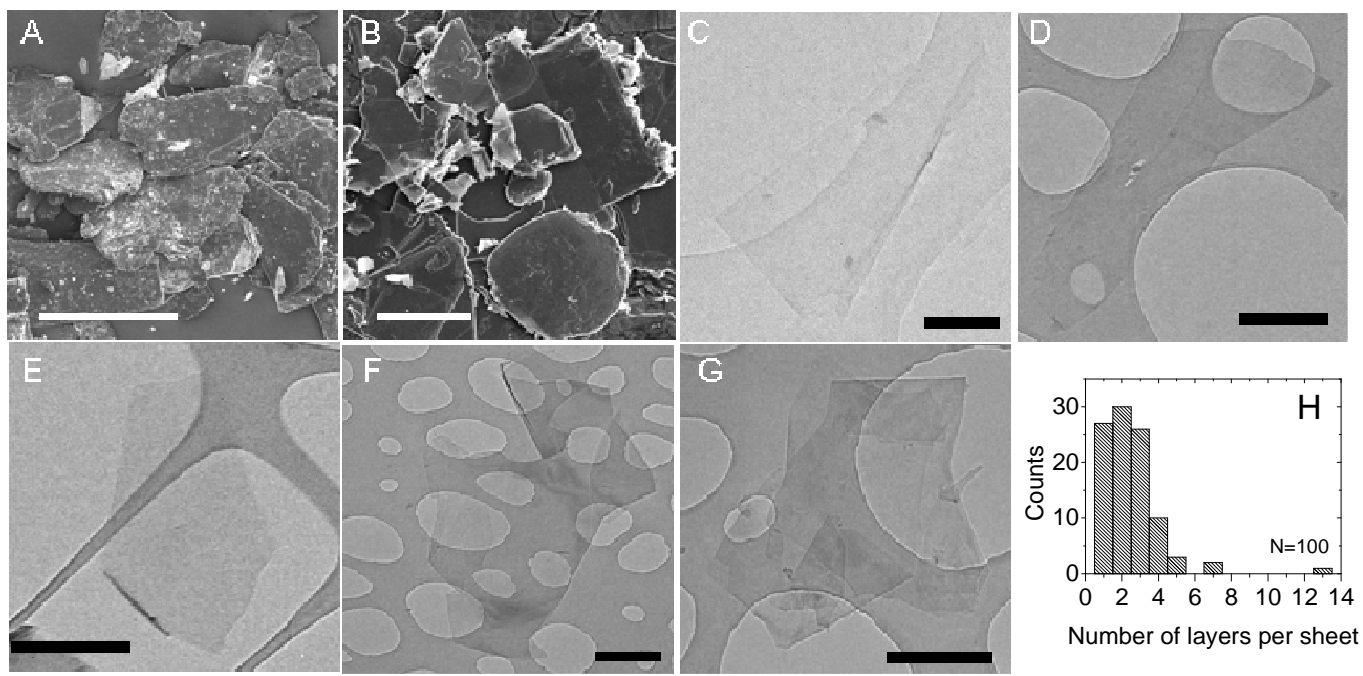

Figure 2 A) SEM image of sieved, pristine graphite (scale: 500μm). B) SEM image of sediment after centrifugation (scale: 25μm). C), D) and E Bright field TEM images of single layer graphene flakes deposited from GBL, DMEU and NMP respectively (scale: 500nm). F) A folded graphene sheet (bright field, deposited from NMP). G) Multi-layer graphene (bright field, deposited from NMP) (scale: 500nm). H) A histogram of the number of visual observations of flakes as a function of the number of monolayers per flake for NMP dispersions. This data allows the calculation of the number fraction of monolayers (28%), the mass fraction of monolayers (12%) and the overall yield of graphene (0.8%) (see table S2 for details).

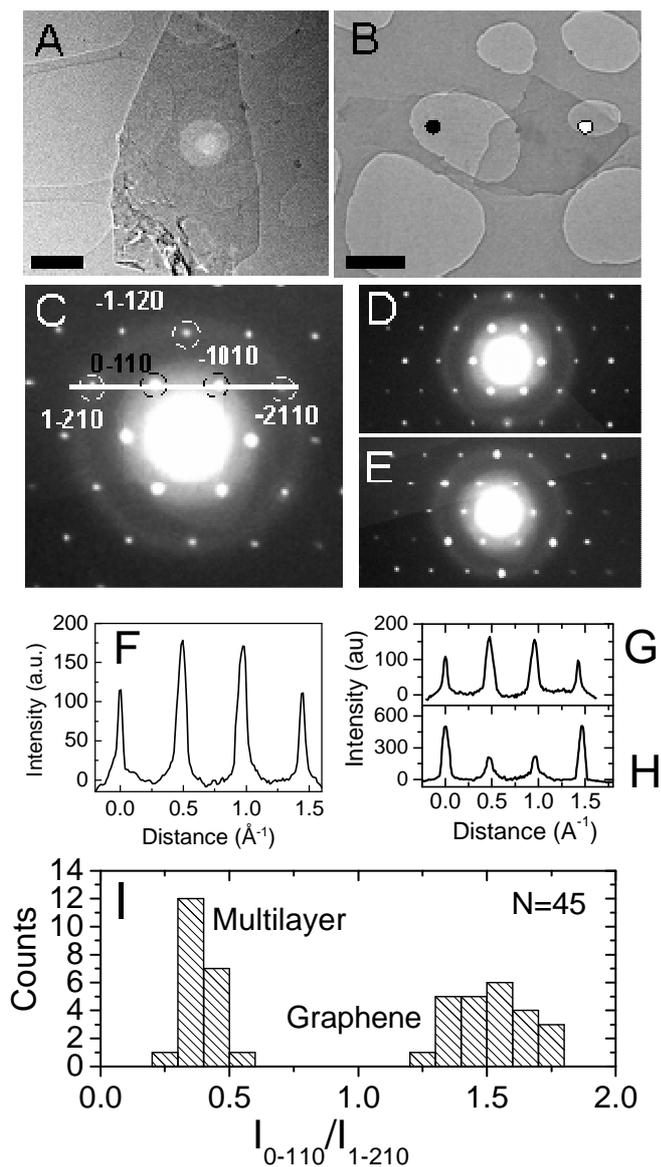

Figure 3 A) and B) HRTEM images of solution cast monolayer and bi-layer graphene respectively. C) Electron diffraction pattern of the sheet in A). The peaks are labelled using Miller-Bravais indices. The same labels also apply to the patterns in D) and E). D) and E) Electron diffraction patterns of the sheet shown in B). The pattern in D) was taken from the position marked with the black spot where the sheet is clearly one layer thick. The pattern in E) was taken from the position marked with the white spot where the sheet is clearly two layers thick. F), G) and H) Diffracted intensity taken along the 1-210 to -2110 axis for the patterns shown in C), D) and E) respectively. I) Histogram of the ratios of the intensity of the {1100} and {2110} diffraction peaks for all the diffraction patterns collected. A ratio greater than 1 is a signature of graphene.